\documentclass[12pt]{article}
\usepackage{amsmath}
\usepackage{amssymb}
\usepackage{amscd}

\numberwithin{equation}{section}

\title{
REVISITING (QUASI-)EXACTLY SOLVABLE  RATIONAL EXTENSIONS OF THE MORSE POTENTIAL}
\author{C. QUESNE\\
{\small \sl Physique Nucl\'eaire Th\'eorique et Physique Math\'ematique,}\\ 
{\small \sl Universit\'e Libre de Bruxelles, Campus de la Plaine CP229,} \\ 
{\small \sl Boulevard~du Triomphe, B-1050 Brussels, Belgium} \\
{\small \sl cquesne@ulb.ac.be}}
\date{ }
\begin{document}
\baselineskip=22pt plus 1pt minus 1pt
\maketitle

\begin{abstract} 
The construction of rationally-extended Morse potentials is analyzed in the framework of first-order supersymmetric quantum mechanics. The known family of extended potentials $V_{A,B,{\rm ext}}(x)$, obtained from a conventional Morse potential $V_{A-1,B}(x)$ by the addition of a bound state below the spectrum of the latter, is re-obtained. More importantly, the existence of another family of extended potentials, strictly isospectral to $V_{A+1,B}(x)$, is pointed out for a well-chosen range of parameter values. Although not shape invariant, such extended potentials exhibit a kind of `enlarged' shape invariance property, in the sense that their partner, obtained by translating both the parameter $A$ and the degree $m$ of the polynomial arising in the denominator, belongs to the same family of extended potentials. The point canonical transformation connecting the radial oscillator to the Morse potential is also applied to exactly solvable rationally-extended radial oscillator potentials to build quasi-exactly solvable rationally-extended Morse ones. 
\end{abstract}

\noindent
Running head: Rational Extensions of Morse Potential


%
%
\newpage

\section{Introduction}

Extending the Morse potential by the addition of some nonsingular function, in such a way that the resulting potential remains exactly solvable (ES), has a long history and has been done along several paths.\par
%
%
In the context of unbroken supersymmetric quantum mechanics (SUSYQM) \cite{cooper}, an extended potential has been built in terms of the confluent hypergeometric function and its derivative and shown to be ES under certain conditions on the potential parameters \cite{junker}. For such a reason, the new potential has been termed `conditionally exactly solvable' \cite{dutra, dutt}.\par
%
%
In the framework of backward Darboux transformations \cite{darboux} (equivalent to the previous approach of unbroken SUSYQM) of (translationally) shape invariant (SI) potentials \cite{gendenshtein}, algebraic deformations of the Morse potential have been considered \cite{gomez04}. These deformations are characterized by the fact that the superpotential is a rational function or a composition of a rational function with an exponential. Special attention has been devoted to the polynomials appearing in the partner wavefunctions and to their properties.\par
%
%
Another method has employed ground- or excited-state wavefunctions of the Morse potential to construct nonsingular isospectral potentials \cite{berger, dutta} by resorting to the well-known nonuniqueness of factorization \cite{mielnik}. The latter indeed allows one to avoid the singularities arising from the use of excited-state wavefunctions in standard SUSYQM \cite{cooper}.\par
%
%
After the introduction of the first families of exceptional orthogonal polynomials (EOP) in the context of Sturm-Liouville theory \cite{gomez10a, gomez09}, the realization of their usefulness in constructing new SI extensions of ES potentials in quantum mechanics \cite{cq08a, bagchi09a, cq09}, and the rapid developments that followed in this area \cite{odake09, odake10a, odake10b, ho11a, gomez10b, gomez11a, sasaki, grandati11a, ho11b, gomez12, odake11, cq11a, cq11b, grandati11b, cq11c}, it soon appeared that only some of the well-known SI potentials led to rational extensions connected with EOP. In this category, one finds the radial oscillator \cite{cq08a, cq09, odake09, odake10a, odake10b, sasaki, grandati11a, ho11b}, the Scarf I (also called trigonometric P\"oschl-Teller or P\"oschl-Teller I) \cite{cq08a, cq09, odake09, odake10a, sasaki, ho11b}, and the generalized P\"oschl-Teller (also termed hyperbolic P\"oschl-Teller or P\"oschl-Teller II) \cite{bagchi09a, odake09, odake10a}.\par
%
%
In such a context, the Morse potential has been recently re-examined by making use of two different (but equivalent) approaches, the Darboux-B\"acklund transformation \cite{grandati11c} and the prepotential method \cite{ho11c}. Both studies have recovered the previously found algebraic deformations \cite{gomez04}.\par
%
%
The purpose of the present work is twofold. First, we will review the construction of rationally-extended Morse potentials by using a standard SUSYQM approach, similar to that previously employed for some other potentials \cite{bagchi09a, cq09, cq11b, cq11c}. This will enable us to point out the existence of a whole family of strictly isospectral extensions for some range of parameter values. Second, we will generalize to rationally-extended potentials the point canonical transformation (PCT) known to connect the radial oscillator to the Morse potential \cite{cooper, haymaker, de, cq08b}. This will lead us to quasi-exactly solvable (QES) \cite{turbiner, shifman, ushveridze} rational extensions.\par
%
%
\section{Rationally-Extended Morse Potentials in First-Order SUSYQM}

\subsection{General results}

As well known, the Morse potential
\begin{equation}
  V_{A,B}(x) = B^2 e^{-2x} - B (2A+1) e^{-x}, \qquad - \infty < x < \infty,  \label{eq:Morse-pot}
\end{equation}
has a minimum $V_{A,B}(x_{\rm min}) = - \frac{1}{4} (2A+1)^2$, for $x_{\rm min}$ such that $e^{-x_{\rm min}} = (2A+1)/(2B)$, provided this quantity is positive. It then has a finite number of bound states with energy \cite{cooper}
\begin{equation}
  \epsilon^{(A)}_{\nu} = - (A-\nu)^2, \qquad \nu = 0, 1, \ldots, \nu_{\rm max}, \qquad A-1 \le \nu_{\rm max} <
  A,  \label{eq:Morse-spectrum}
\end{equation}
provided $A > 0$ (hence $B > 0$ too).\footnote{In this paper, we take units wherein $\hbar = 2m = 1$.} The corresponding bound-state wavefunctions can be expressed in terms of (generalized) Laguerre polynomials as
\begin{equation}
  \varphi^{(A)}_{\nu}(x) \propto \exp[- (A-\nu) x - B e^{-x}] L^{(2A-2\nu)}_{\nu}(2B e^{-x}) \propto \xi_A(z)
  L^{(2A-2\nu)}_{\nu}(z),  \label{eq:Morse-wf}
\end{equation}
with
\begin{equation}
  z = 2B e^{-x}, \qquad \xi_A(z) = z^{A-\nu} e^{- \frac{1}{2} z}, \qquad 0 < z < \infty.  \label{eq:z}
\end{equation}
\par
%
%
In first-order SUSYQM \cite{cooper}, one considers a pair of SUSY partners
\begin{equation}
\begin{split}
  H^{(+)} & = \hat{A}^{\dagger} \hat{A} = - \frac{d^2}{dx^2} + V^{(+)}(x) - \epsilon, \\
  H^{(-)} & = \hat{A} \hat{A}^{\dagger} = - \frac{d^2}{dx^2} + V^{(-)}(x) - \epsilon, \\
  \hat{A}^{\dagger} & = - \frac{d}{dx} + W(x), \\
  \hat{A} & = \frac{d}{dx} + W(x), \\ 
  V^{(\pm)}(x) & = W^2(x) \mp W'(x) + \epsilon ,  
\end{split}  \label{eq:SUSY}
\end{equation}
which intertwine with the first-order differential operators $\hat{A}$ and $\hat{A}^{\dagger}$ as $\hat{A} H^{(+)} = H^{(-)} \hat{A}$ and $\hat{A}^{\dagger} H^{(-)} = H^{(+)} \hat{A}^{\dagger}$. Here $W(x)$ is the superpotential, which can be expressed as $W(x) = - \bigl(\log \phi(x)\bigr)'$ in terms of a (nodeless) seed solution $\phi(x)$ of the initial Schr\"odinger equation
\begin{equation}
  \left(- \frac{d^2}{dx^2} + V^{(+)}(x)\right) \phi(x) = \epsilon \phi(x),  \label{eq:SE}
\end{equation}
$\epsilon$ is the factorization energy, assumed smaller than or equal to the ground-state energy $\epsilon^{(+)}_0$ of $V^{(+)}(x)$, and a prime denotes a derivative with respect to $x$. For $\epsilon = \epsilon^{(+)}_0$ and $\phi(x) = \varphi^{(+)}_0(x)$ corresponding to the ground state of $V^{(+)}(x)$, the partner potential $V^{(-)}(x)$ has the same bound-state spectrum as $V^{(+)}(x)$, except for the ground-state energy which is removed (case {\it i}). For $\epsilon < \epsilon^{(+)}_0$, in which case $\phi(x)$ is a nonnormalizable function, $V^{(-)}(x)$ has the same spectrum as $V^{(+)}(x)$ if $\phi^{-1}(x)$ is also nonnormalizable (case {\it ii} or isospectral case) or it has an extra bound-state energy $\epsilon$ below $\epsilon^{(+)}_0$, corresponding to the wavefunction $\phi^{-1}(x)$, if the latter is normalizable (case {\it iii}).\par
%
%
{}For $V^{(+)}(x) = V_{A,B}(x)$, it is well known \cite{cooper} that $\epsilon = \epsilon^{(+)}_0 = \epsilon^{(A)}_0$ and $\phi(x) = \varphi^{(+)}_0(x) = \varphi^{(A)}_0(x)$ lead to $V^{(-)}(x) = V_{A-1,B}(x)$, showing that the Morse potential is SI \cite{gendenshtein}.\par
%
%
To construct rational extensions of the Morse potential, we have to determine all well-behaved (i.e., nodeless) solutions $\phi(x)$ of (\ref{eq:SE}) with $\epsilon < \epsilon^{(A)}_0 = - A^2$ that are of polynomial type. For such a purpose, let us use the ansatz
\begin{equation}
  \phi\bigl(x(z)\bigr) = z^{\lambda} e^{- \frac{1}{2} z} f(z)
\end{equation}
in the Schr\"odinger equation (\ref{eq:SE}), rewritten in the variable $z$ defined in (\ref{eq:z}),
\begin{equation}
  \left[- z^2 \frac{d^2}{dz^2} - z \frac{d}{dz} + \frac{1}{4} z^2 - \left(A + \frac{1}{2}\right) z - \epsilon\right]
  \phi\bigl(x(z)\bigr) = 0.
\end{equation}
Here $\lambda$ and $f(z)$ denote some constant and some function, respectively. Provided
\begin{equation}
  \epsilon = - \lambda^2, \qquad a = \lambda - A, \qquad b = 2\lambda + 1,  \label{eq:conditions}
\end{equation}
the resulting equation for $f(z)$ reduces to the confluent hypergeometric equation
\begin{equation}
  \left[z \frac{d^2}{dz^2} + (b-z) \frac{d}{dz} - a \right] f(z) = 0,  \label{eq:confluent}
\end{equation}
whose regular solution is ${}_1F_1(a; b; z)$. Equation (\ref{eq:confluent}) admits four polynomial-type solutions, expressed in terms of Laguerre polynomials, if and only if either $a$ or $b-a$ is an integer \cite{erdelyi} (see also Ref.\ \cite{gomez04}),
\begin{equation}
\begin{split}
  f_1(z) & = {}_1F_1(a; b; z) \propto L^{(b-1)}_m(z) \qquad \text{for\ } a = - m,  \\
  f_2(z) & = z^{1-b} {}_1F_1(a-b+1; 2-b; z) \propto z^{1-b} L^{(1-b)}_m(z) \qquad \text{for\ } b-a = m+1, \\
  f_3(z) & = e^z {}_1F_1(b-a; b; -z) \propto e^z L^{(b-1)}_m(-z) \qquad \text{for\ } b-a = - m,  \\
  f_4(z) & = z^{1-b} e^z {}_1F_1(1-a; 2-b; -z) \propto z^{1-b} e^z L^{(1-b)}_m(-z) \qquad \text{for\ } a = m+1.      
\end{split}  \label{eq:f}
\end{equation}
\par
%
%
It remains to combine Eq.\ (\ref{eq:conditions}) with the condition found for $a$ or $b-a$ in (\ref{eq:f}) and to look for those cases where $\epsilon < - A^2$ and the Laguerre polynomial has no zero for $z \in (0, \infty)$. To check the last condition, we use Kienast-Lawton-Hahn's theorem on the zeros of Laguerre polynomials \cite{erdelyi} (see also Ref.\ \cite{grandati11a}). From $f_1(z)$ and $f_3(z)$ (or, equivalently, from $f_2(z)$ and $f_4(z)$), we arrive at the following two acceptable factorization functions
\begin{eqnarray}
  \phi^{\rm II}_{A,m}(x) & = & \chi^{\rm II}_{A,m}(z) L^{(2A-2m)}_m(z) \nonumber \\
  & \propto & \exp[- (A-m)x - Be^{-x}] L^{(2A-2m)}_m(2Be^{-x}) \nonumber \\
  && \text{if\ } m = 1, 2, 3, \ldots \text{\ and\ } A < \frac{m}{2},  \label{eq:phi-II}
\end{eqnarray}
\begin{eqnarray}
  \phi^{\rm III}_{A,m}(x) & = & \chi^{\rm III}_{A,m}(z) L^{(-2A-2m-2)}_m(-z) \nonumber \\ 
  & \propto & \exp[(A+m+1)x + Be^{-x}] L^{(-2A-2m-2)}_m(-2Be^{-x}) \nonumber \\
  && \text{if\ } m = 2, 4, 6, \ldots,  \label{eq:phi-III}
\end{eqnarray}
with 
\begin{equation}
  \chi^{\rm II}_{A,m}(z) = z^{A-m} e^{- \frac{1}{2} z}, \qquad \chi^{\rm III}_{A,m}(z) = z^{-A-m-1} 
  e^{\frac{1}{2} z},  \label{eq:chi} 
\end{equation}
and corresponding energies $\epsilon^{\rm II}_{A,m} = - (A-m)^2$, $\epsilon^{\rm III}_{A,m} = - (A+m+1)^2$, respectively. The inverse of $\phi^{\rm III}_{A,m}(x)$ is normalizable, in contrast with that of $\phi^{\rm II}_{A,m}(x)$.\par
%
%
A superscript II or III has been introduced in Eqs.\ (\ref{eq:phi-II}), (\ref{eq:phi-III}), and (\ref{eq:chi}), by analogy with what is often done in the case of extended radial oscillator potentials pertaining to the L2 and L3 series, respectively \cite{grandati11a, cq11c}. The former series is associated with Laguerre polynomials with negative argument and positive variable, while for the latter both the argument and the variable are negative. For the extended Morse potentials, no counterpart of the L1 series, corresponding to positive argument and negative variable, is obtained. It is worth stressing that $\phi^{\rm III}_{A,m}(x)$ has been first derived in Ref.\ \cite{gomez04}, then reconsidered in Refs.\ \cite{grandati11c} and \cite{ho11c}, but that to the best of the author's knowledge, $\phi^{\rm II}_{A,m}(x)$ has not been mentioned so far.\par
%
%
To obtain some rationally-extended Morse potentials $V_{A,B,{\rm ext}}(x)$ with given $A$ and $B$, we have to start from a conventional Morse potential $V_{A',B}(x)$ with some different $A'$, but the same $B$ (hence $z$ remains unchanged). From Eqs.\ (\ref{eq:SUSY}), (\ref{eq:phi-II}), (\ref{eq:phi-III}), and (\ref{eq:chi}), it is straightforward to get
\begin{equation}
\begin{split}
  & V^{(+)}(x) = V_{A',B}(x), \qquad V^{(-)}(x) = V_{A,B,{\rm ext}}(x) = V_{A,B}(x) + V_{A,B,{\rm rat}}(x), \\
  & V_{A,B,{\rm rat}}(x) = - 2z \biggl\{\frac{\dot{g}^{(A)}_m}{g^{(A)}_m} + z 
        \biggl[\frac{\ddot{g}^{(A)}_m}{g^{(A)}_m} - \biggl(\frac{\dot{g}^{(A)}_m}
        {g^{(A)}_m}\biggr)^2\biggr]\biggr\}, 
\end{split}  \label{eq:partners}  
\end{equation}
where a dot denotes a derivative with respect to $z$. According to the choice made for the factorization function $\phi(x)$, we may distinguish the two cases
\begin{eqnarray}
  & ({\rm II}) \; &  A' = A+1, \quad \phi = \phi^{\rm II}_{A+1,m}, \quad g^{(A)}_m(z) = L^{(2A+2-2m)}_m(z),
         \nonumber \\
  && m = 1, 2, 3, \ldots, \quad -1 < A < \frac{m-2}{2}; \label{eq:type-II} \\
  & ({\rm III}) \; &  A' = A-1, \quad \phi = \phi^{\rm III}_{A-1,m}, \quad g^{(A)}_m(z) = L^{(-2A-2m)}_m(-z), 
         \nonumber \\
  && m = 2, 4, 6, \ldots, \quad A>1.  \label{eq:type-III}
\end{eqnarray}
Note that $B>0$ everywhere.\par
%
%
\subsection{Type II rationally-extended Morse potentials}

In type II case, $V^{(+)}(x)$ and $V^{(-)}(x)$ are isospectral (case {\it ii} of SUSYQM) and their common bound-state spectrum is given by
\begin{equation}
  \epsilon^{(+)}_{\nu} = \epsilon^{(-)}_{\nu} = - (A + 1 - \nu)^2, \qquad \nu = 0, 1, \ldots, \nu_{\rm max},
  \qquad A \le \nu_{\rm max} < A+1.
\end{equation}
When $A$ varies in the range $-1 < A < \frac{m-2}{2}$, the number of bound states $\nu_{\rm max} + 1$ goes from one to $\left[\frac{m+1}{2}\right]$.\par
%
%
{}From the bound-state wavefunctions $\varphi^{(+)}_{\nu}(x) \propto \xi_{A+1}(z) L^{(2A+2-2\nu)}_{\nu}(z)$, $\nu = 0$, 1, \ldots, $\nu_{\rm max}$, of $V^{(+)}(x)$, those of $V^{(-)}(x)$ are obtained as
\begin{equation}
  \varphi^{(-)}_{\nu}(x) \propto \hat{A} \varphi^{(+)}_{\nu}(x) \propto \frac{\xi_{A+1}(z)}{g^{(A)}_m(z)}
  y^{(A)}_n(z), \qquad \nu = 0, 1, \ldots, \nu_{\rm max},  \label{eq:partner-wf}
\end{equation}
with
\begin{equation}
  \hat{A} = - z \left(\frac{d}{dz} + \frac{1}{2} - \frac{A+1-m}{z} - \frac{\dot{g}^{(A)}_m}{g^{(A)}_m}\right).
\end{equation}
In (\ref{eq:partner-wf}), $y^{(A)}_n(z)$ denotes some $n$th-degree polynomial in $z$, defined by
\begin{equation}
  y^{(A)}_n(z) = \left[g^{(A)}_m \left(- z \frac{d}{dz} + \nu - m\right) + z \dot{g}^{(A)}_m\right]
  L^{(2A+2-2\nu)}_{\nu}(z).  \label{eq:def-y}
\end{equation}
This definition seems to imply that $n = m + \nu$. Nevertheless, from the relation
\begin{equation}
  y^{(A)}_n(z) = (2A + 2 - \nu) g^{(A)}_m(z) L^{(2A+2-2\nu)}_{\nu-1}(z) - (2A + 2 - m) g^{(A-1)}_{m-1}(z)
  L^{(2A+2-2\nu)}_{\nu}(z),
\end{equation}
directly obtainable from the right-hand side of (\ref{eq:def-y}) and some elementary properties of Laguerre polynomials \cite{gradshteyn}, we actually deduce that
\begin{equation}
  n = m + \nu - 1.
\end{equation}
Note that with the normalization assumed in (\ref{eq:def-y}), the highest-degree term of $y^{(A)}_{m+\nu-1}(z)$ is given by $(m-\nu) (m+\nu-2A-2) (-z)^{m+\nu-1}/(m!\, \nu!)$. As a special case, the ground-state wavefunction of $V^{(-)}(x)$ can be written as
\begin{equation}
  \varphi^{(-)}_0(x) \propto \frac{\xi_{A+1}(z)}{g^{(A)}_m(z)} g^{(A-1)}_{m-1}(z). \label{eq:gs-wf}
\end{equation}
\par
%
%
On the other hand, by directly inserting Eq.\ (\ref{eq:partner-wf}) in the Schr\"odinger equation for $V^{(-)}(x)$, we arrive at the following second-order differential equation for $y^{(A)}_{m+\nu-1}(z)$,
\begin{eqnarray}
  && \biggl\{z \frac{d^2}{dz^2} + \biggl[2A+3-2\nu - z\biggl(1 + 2\frac{\dot{g}^{(A)}_m}{g^{(A)}_m}\biggr)
       \biggr] \frac{d}{dz} - 2 (2A+2-m-\nu-z) \frac{\dot{g}^{(A)}_m}{g^{(A)}_m} \nonumber \\
  && \quad - (m-\nu+1)\biggr\} y^{(A)}_{m+\nu-1}(z) = 0, \qquad \nu = 0, 1, \ldots, \nu_{\rm max}.
       \label{eq:diff-eq}
\end{eqnarray}
\par
%
%
Let us illustrate the results obtained here by considering the $m=1$, 2, and 3 special cases. The rational part of the extended potentials can be written as
\begin{equation}
  V_{A,B,{\rm rat}}(x) = \frac{N_1(x)}{D(x)} + \frac{N_2(x)}{D^2(x)},  \label{eq:V-rat}
\end{equation}
where
\begin{equation}
\begin{split}
  N_1(x) & = 2 (2A+1), \\
  N_2(x) & = 2 (2A+1)^2, \\
  D(x) & = 2Be^{-x} - 2A - 1,
\end{split}
\end{equation}
with $-1 < A < -\frac{1}{2}$ for $m=1$,
\begin{equation}
\begin{split}
  N_1(x) & = 8A \bigl(Be^{-x} + 1\bigr), \\
  N_2(x) & = 8A^2 \bigl(4Be^{-x} - 2A + 1\bigr), \\
  D(x) & = 2B^2 e^{-2x} - 4ABe^{-x} + A (2A-1),
\end{split}  \label{eq:example-II}
\end{equation}
with $-1 < A < 0$ for $m=2$, and
\begin{equation}
\begin{split}
  N_1(x) & = 3 (2A-1) \bigl[4B^2 e^{-2x} - 2 (2A-5) Be^{-x} + 3 (2A+1)\bigr], \\
  N_2(x) & = 9 (2A-1)^2 \bigl[2 (2A+7) B^2 e^{-2x} - 4 (A-1)(2A+3) Be^{-x} \\
  & \quad + (A-1)(2A-3)(2A+1)\bigr], \\
  D(x) & = 4B^3 e^{-3x} - 6(2A-1) B^2 e^{-2x} + 6(A-1)(2A-1) Be^{-x} \\
  & \quad - (A-1)(2A-1)(2A-3),
\end{split}
\end{equation}
with $-1 < A < \frac{1}{2}$ for $m=3$. In the first two cases, there is a single bound state with energy $\epsilon^{(-)}_0 = - (A+1)^2$ and wavefunction
\begin{equation}
  \varphi^{(-)}_0(x) \propto \frac{\exp\bigl[- (A+1)x - Be^{-x}\bigr]}{D(x)}
\end{equation}
or
\begin{equation}
  \varphi^{(-)}_0(x) \propto \frac{\exp\bigl[- (A+1)x - Be^{-x}\bigr]}{D(x)} \bigl(2Be^{-x} - 2A + 1\bigr),
\end{equation}
respectively. In contrast, in the third case, there may be up to two bound states
\begin{equation}
  \varphi^{(-)}_0(x) \propto \frac{\exp\bigl[- (A+1)x - Be^{-x}\bigr]}{D(x)} \bigl[2B^2 e^{-2x} - 4(A-1) Be^{-x}
  + (A-1)(2A-3)\bigr]
\end{equation}
and
\begin{equation}
\begin{split}
  \varphi^{(-)}_1(x) & \propto \frac{\exp\bigl[- (A+1)x - Be^{-x}\bigr]}{D(x)} \bigl[8B^3 e^{-3x} - 12(2A-1) B^2
      e^{-2x} \\
  & \quad + 6(2A-1)^2 Be^{-x} - (2A+1)(2A-1)(2A-3)\bigr],
\end{split}
\end{equation}
corresponding to $\epsilon^{(-)}_0 = - (A+1)^2$ and $\epsilon^{(-)}_1 = - A^2$, provided $0 < A < \frac{1}{2}$. Only the first one of them, however, exists for $-1 < A \le 0$.\par
%
%
It is worth observing that the addition of $V_{A,B,{\rm rat}}(x)$ to $V_{A,B}(x)$ has the effect of increasing the number of bound states by one, since for the allowed values of parameter $A$, the core part of the extended potential has no bound state for $m=1$ nor for $m=2$, and may have zero or one bound state for $m=3$ according to whether $-1 < A \le 0$ or $0 < A < \frac{1}{2}$. This observation remains true for higher $m$ values.\par
%
%
\subsection{Type III rationally-extended Morse potentials}

In type III case, $V^{(+)}(x)$ and $V^{(-)}(x)$ are not isospectral anymore (case {\it iii} of SUSYQM). Their bound-state spectra are given instead by
\begin{equation}
  \epsilon^{(+)}_{\nu} = - (A-1-\nu)^2, \quad \nu = 0, 1, \ldots, \nu_{\rm max}, \quad A-2 \le \nu_{\rm max} 
      < A-1,
\end{equation}
and
\begin{equation} 
  \epsilon^{(-)}_{\nu} = - (A-1-\nu)^2, \quad \nu = -m-1, 0, 1, \ldots, \nu_{\rm max}, \quad A-2 \le 
      \nu_{\rm max} < A-1,      
\end{equation}
the ground state of $V^{(-)}(x)$ corresponding to $\epsilon^{(-)}_{-m-1} = \epsilon^{\rm III}_{A-1,m} = - (A+m)^2$.\par
%
%
The bound-state wavefunctions of $V^{(-)}(x)$ can be written as
\begin{equation}
  \varphi^{(-)}_{\nu}(x) \propto \frac{\xi_{A-1}(z)}{g^{(A)}_m(z)} y^{(A)}_n(z), \qquad n = m+\nu+1, \qquad 
  \nu = -m-1, 0, 1, \ldots, \nu_{\rm max},
\end{equation}
where $y^{(A)}_n(z)$ is an $n$-th-degree polynomial in $z$. For the ground state,
\begin{equation}
  \varphi^{(-)}_{-m-1}(x) \propto \left(\phi^{\rm III}_{A-1,m}(x)\right)^{-1}, \qquad y^{(A)}_0(z) = 1,
\end{equation}
while for the excited states
\begin{equation}
  \varphi^{(-)}_{\nu}(x) \propto \hat{A} \varphi^{(+)}_{\nu}(x), \qquad \nu = 0, 1, \ldots, \nu_{\rm max}, 
\end{equation}
with $\varphi^{(+)}_{\nu}(x) \propto \xi_{A-1}(z) L^{(2A-2-2\nu)}_{\nu}(z)$ and
\begin{equation}
  \hat{A} = - z \left(\frac{d}{dz} - \frac{1}{2} + \frac{A+m}{z} - \frac{\dot{g}^{(A)}_m}{g^{(A)}_m}\right).
\end{equation}
For $n = m+1$, $m+2$, \ldots, $m+1+\nu_{\rm max}$, we may therefore define $y^{(A)}_n(z)$ as
\begin{equation}
  y^{(A)}_n(z) = \left[g^{(A)}_m \left(-z \frac{d}{dz} - 2A + 1 - m + \nu + z\right) + z \dot{g}^{(A)}_m\right]
  L^{(2A-2-2\nu)}_{\nu}(z).  \label{eq:def-y-bis}
\end{equation}
Standard properties of Laguerre polynomials \cite{gradshteyn} may be used to rewrite Eq.\ (\ref{eq:def-y-bis}) in either of the two equivalent forms
\begin{equation}
  y^{(A)}_n(z) = (m+1) g^{(A-1)}_{m+1}(z) L^{(2A-2-2\nu)}_{\nu}(z) + (2A-2-\nu) g^{(A)}_m(z)
  L^{(2A-2-2\nu)}_{\nu-1}(z) 
\end{equation}
or
\begin{equation}
  y^{(A)}_n(z) = (2A+m) g^{(A+1)}_{m-1}(z) L^{(2A-2-2\nu)}_{\nu}(z) - (\nu+1) g^{(A)}_m(z)
  L^{(2A-2-2\nu)}_{\nu+1}(z). 
\end{equation}
The latter expression was already given in previous studies \cite{grandati11c, ho11c}, but the former is a new one, which has the advantage of directly providing us with the polynomial appearing in the first-excited state wavefunction,
\begin{equation}
  y^{(A)}_{m+1}(z) = (m+1) g^{(A-1)}_{m+1}(z) = (m+1) L^{(-2A-2m)}_{m+1}(-z).
\end{equation}
We may also observe that with the normalization chosen in (\ref{eq:def-y-bis}), the highest-degree term of $y^{(A)}_{m+\nu+1}(z)$ is given by $(-1)^{\nu} z^{m+\nu+1}/(m!\, \nu!)$ for $\nu = 0$, 1, \ldots, $\nu_{\rm max}$.\par
%
%
{}Finally, for type III polynomials, the counterpart of the second-order differential equation (\ref{eq:diff-eq}) reads
\begin{eqnarray}
  && \biggl\{z \frac{d^2}{dz^2} + \biggl[2A-1-2\nu - z\biggl(1 + 2\frac{\dot{g}^{(A)}_m}{g^{(A)}_m}\biggr)
       \biggr] \frac{d}{dz} + (m+\nu+1) \biggl(1 + 2\frac{\dot{g}^{(A)}_m}{g^{(A)}_m}\biggr)\biggr\} 
       \nonumber \\
  && \quad \times y^{(A)}_{m+\nu+1}(z) = 0, \qquad \nu = -m-1, 0, 1, \ldots, \nu_{\rm max}.
\end{eqnarray}
\par
%
%
The results obtained here may be illustrated by considering the lowest allowed $m$ value, namely $m=2$. In such a case, the rational part of the extended potential takes the form (\ref{eq:V-rat}) with
\begin{equation}
\begin{split}
  N_1(x) & = 8 (A+1) \bigl(Be^{-x} - 1\bigr), \\
  N_2(x) & = - 8 (A+1)^2 \bigl(4Be^{-x} - 2A - 3\bigr), \\
  D(x) & = 2B^2 e^{-2x} - 4 (A+1) Be^{-x} + (A+1) (2A+3),
\end{split}  \label{eq:example-III}
\end{equation}
where $A>1$. Equation (\ref{eq:example-III}) may be compared with Eq.\ (\ref{eq:example-II}), corresponding to the other quadratic-type extended potential.\par
%
%
\subsection{Partner of rationally-extended Morse potentials in unbroken SUSYQM}

If we take any of the rationally-extended Morse potentials obtained so far as the starting potential $\bar{V}^{(+)}(x) = V^{(-)}(x)$ in first-order SUSYQM, it is interesting to determine its partner $\bar{V}^{(-)}(x)$ when its ground state is deleted (case {\sl i} of SUSYQM). In type III case, this is of course the inverse transformation of that carried out in Secs.\ 2.1 and 2.3, so that $\bar{V}^{(-)}(x) = V^{(+)}(x)$. Hence it only remains to consider type II extended potentials.\par
%
%
In such a case, the new factorization function corresponds to the ground-state wavefunction (\ref{eq:gs-wf}) of $V^{(-)}(x)$, which leads to the new superpotential
\begin{equation}
  \bar{W}(x) = A+1 - \frac{1}{2}z + z\left(\frac{\dot{g}^{(A-1)}_{m-1}}{g^{(A-1)}_{m-1}} - 
  \frac{\dot{g}^{(A)}_m}{g^{(A)}_m}\right),
\end{equation}
written in terms of the variable $z$. The searched for partner is then given by
\begin{equation}
  \bar{V}^{(-)}(x) = \bar{V}^{(+)}(x) + 2 \bar{W}'(x)
\end{equation}
with $\bar{V}^{(+)}(x) = V^{(-)}(x)$ expressed in terms of $g^{(A)}_m(z)$ and its derivatives as in (\ref{eq:partners}). A straightforward calculation yields
\begin{equation}
  \bar{V}^{(-)}(x) = V_{A-1,B}(x) - 2z \biggl\{\frac{\dot{g}^{(A-1)}_{m-1}}{g^{(A-1)}_{m-1}} + z 
        \biggl[\frac{\ddot{g}^{(A-1)}_{m-1}}{g^{(A-1)}_{m-1}} - \biggl(\frac{\dot{g}^{(A-1)}_{m-1}}
        {g^{(A-1)}_{m-1}}\biggr)^2\biggr]\biggr\}.
\end{equation}
\par
%
%
It is remarkable that although $V_{A,B,{\rm ext}}(x)$ is not translationally SI, its partner belongs to an enlarged family of extended Morse potentials, wherein both the potential parameter $A$ and the polynomial degree $m$ are translated: $A \to A-1$, $m \to m-1$. In other words, what we have done here is to go from $\bar{V}^{(+)}(x) = V^{(m)}_{A,B,{\rm ext}}(x)$ to $\bar{V}^{(-)}(x) = V^{(m-1)}_{A-1,B,{\rm ext}}(x)$, where we have appended a superscript to specify the polynomial degree. The final potential $\bar{V}^{(-)}(x)$ having one bound state less than the initial one $\bar{V}^{(+)}(x)$, its spectrum is given by $- (A-\nu)^2$, $\nu=0$, 1, \ldots, $\nu_{\rm max}-1$ ($A \le \nu_{\rm max} < A+1$). As a consequence, for $m=1$ and $m=2$, we arrive at a conventional Morse potential and an extended Morse one with no bound state, respectively. It is only from $m=3$ upwards that extended Morse potentials with at least one bound state may be obtained.\par
%
%
Putting together the first step from $V^{(+)}(x)$ to $V^{(-)}(x)$ and the second one from $\bar{V}^{(+)}(x) = V^{(-)}(x)$ to $\bar{V}^{(-)}(x)$ (see Ref.\ \cite{bagchi09a} and references quoted therein), we arrive at a reducible second-order SUSYQM transformation from a conventional Morse potential $V_{A+1,B}(x)$ to an extended one $V^{(m-1)}_{A-1,B,{\rm ext}}(x)$. Since going from $V_{A+1,B}(x)$ to $V^{(m-1)}_{A-1,B,{\rm ext}}(x)$ can be achieved along another path by combining the usual unbroken SUSYQM transformation relating the two conventional Morse potentials $V_{A+1,B}(x)$ and $V_{A,B}(x)$ \cite{cooper} with the broken one connecting $V_{A,B}(x)$ to $V^{(m-1)}_{A-1,B,{\rm ext}}(x)$ (obtained by substituting $A-1$ and $m-1$ for $A$ and $m$ in Secs.\ 2.1 and 2.2), we finally get the following commutative diagram:
\begin{equation}
\begin{CD}
  V_{A+1,B}(x) @>\text{unbroken}>> V_{A,B}(x)\\
  @V\text{broken}VV @VV\text{broken}V\\
  V_{A,B,{\rm ext}}^{(m)}(x) @>>\text{unbroken}> V_{A-1,B,{\rm ext}}^{(m-1)}(x)
\end{CD}
\end{equation}
This is another example of the possible existence of different intermediate Hamiltonians in higher-order SUSYQM \cite{cq11a} or in type A $\cal N$-fold supersymmetry \cite{bagchi09b}.\par
%
%
\section{Point Canonical Transformation Relating Rationally-Extended Radial Oscillator and Morse Potentials}

\subsection{Going from the radial oscillator to the Morse potential}

To start with, let us briefly review the case of conventional potentials and consider the Schr\"odinger equation for a radial oscillator potential
\begin{equation}
  \left(- \frac{d}{dr^2} + V_l(r)\right) \psi^{(l)}_{\nu}(r) = E^{(l)}_{\nu} \psi^{(l)}_{\nu}(r), \qquad 0 < r < \infty,
  \label{eq:SE-RO}
\end{equation}
where
\begin{equation}
\begin{split}
  V_l(r) & = \frac{1}{4} \omega^2 r^2 + \frac{l(l+1)}{r^2}, \\
  E^{(l)}_{\nu} & = \omega \left(2\nu + l + \frac{3}{2}\right), \qquad \nu=0, 1, 2, \ldots, \\
  \psi^{(l)}_{\nu}(r) & \propto r^{l+1} e^{-\frac{1}{4}\omega r^2} L^{(l+\frac{1}{2})}_{\nu}\left(\frac{1}{2}
       \omega r^2\right) \propto \eta_l(z) L^{(\alpha)}_{\nu}(z), \qquad \nu=0, 1, 2, \ldots,
\end{split}  \label{eq:RO-pot}
\end{equation}
with
\begin{equation}
  z = \frac{1}{2} \omega r^2, \qquad \alpha = l + \frac{1}{2}, \qquad \eta_l(z) = z^{\frac{1}{4}(2\alpha+1)}
  e^{- \frac{1}{2}z}.
\end{equation}
\par
%
%
The changes of variable and of function \cite{cooper, haymaker, de, cq08b}
\begin{equation}
  r = e^{- \frac{1}{2}x}, \qquad \psi^{(l)}_{\nu}(r) = e^{- \frac{1}{4}x} \varphi_{\nu, A_0}(x)  \label{eq:PCT}
\end{equation}
transform Eq.\ (\ref{eq:SE-RO}) into the Schr\"odinger equation for a Morse potential
\begin{equation}
  \left(- \frac{d^2}{dx^2} + V_{A_{\nu},B}(x)\right) \varphi_{\nu,A_0}(x) = \epsilon \varphi_{\nu,A_0}(x),
  \label{eq:SE-Morse-map}
\end{equation}
for some fixed energy $\epsilon$ defined by
\begin{equation}
  \epsilon = - A_0^2 = - \frac{1}{4}\left(l + \frac{1}{2}\right)^2.  \label{eq:epsilon-map}
\end{equation}
Here $V_{A_{\nu},B}(x)$ is given by (\ref{eq:Morse-pot}) with $A$ replaced by the $\nu$-dependent parameter
\begin{equation}
  A_{\nu} = A_0 + \nu, \qquad A_0 = \frac{1}{2}\left(l + \frac{1}{2}\right) = \sqrt{|\epsilon|},
\end{equation}
while
\begin{equation}
  B = \frac{1}{4} \omega
\end{equation}
remains constant. The wavefunction $\varphi_{\nu,A_0}(x)$, corresponding to the energy $\epsilon$, can be obtained by applying (\ref{eq:PCT}) to the radial oscillator wavefunction $\psi^{(l)}_{\nu}(r)$ and is given by\footnote{It should be noted that transformation (\ref{eq:PCT}) results in functions $\varphi_{\nu,A_0}(x)$ that are normalized with respect to an unconventional scalar product \cite{cq08b}.}
\begin{equation}
  \varphi_{\nu,A_0}(x) \propto \exp\bigl(- A_0 x - Be^{-x}\bigr) L^{(2A_0)}_{\nu} (z), \qquad z = 2Be^{-x}.
\end{equation}
\par
%
%
It is important to stress that the Hamiltonian for a single radial oscillator $V_l(r)$, with a given frequency $\omega$ and a given angular momentum quantum number $l$, is transformed into a hierarchy of Hamiltonians of the Morse family, corresponding to $V_{A_{\nu},B}(x)$ with $A_{\nu} = A_0 + \nu$, $\nu=0$, 1, 2,~\ldots, and constant $A_0$, $B$.\par
%
%
Although, for a given $V_{A_{\nu},B}(x)$, we get only a single eigenvalue $\epsilon$ and the corresponding eigenfunction $\varphi_{\nu,A_0}(x)$, it is possible to retrieve the whole Morse spectrum in the following way: on forgetting the map for a moment and focusing on a single Morse potential with given values of $A_{\nu} = \bar{A}$ and $B$, it is obvious that such a potential appears in a finite number of equations of type (\ref{eq:SE-Morse-map}) since, for $A_0$ in (\ref{eq:epsilon-map}), we may choose any of the values $\bar{A} - \bar{\nu}$ with $\bar{\nu} = 0$, 1, \ldots, $\bar{\nu}_{\rm max}$ ($\bar{A}-1 \le \bar{\nu}_{\rm max} < \bar{A}$). Hence, the resulting energy spectrum $- (\bar{A} - \bar{\nu})^2$, $\bar{\nu} = 0$, 1, \ldots, $\bar{\nu}_{\rm max}$, coincides with the standard one $\epsilon^{(\bar{A})}_{\bar{\nu}}$, as given in (\ref{eq:Morse-spectrum}).\par
%
%
\subsection{Going from extended radial oscillators to extended Morse potentials}

In Eq.\ (\ref{eq:SE-RO}), let us now replace the conventional radial oscillator potential by some rationally-extended one. There exist three different types of such extended potentials,
\begin{equation}
\begin{split}
  V_{l,{\rm ext}}(r) & = V_l(r) + V_{l,{\rm rat}}(r), \\
  V_{l,{\rm rat}}(r) & = - 2\omega \biggl\{\frac{\dot{g}^{(\alpha)}_m}{g^{(\alpha)}_m} + 2z 
       \biggl[\frac{\ddot{g}^{(\alpha)}_m}{g^{(\alpha)}_m}
       - \biggl(\frac{\dot{g}^{(\alpha)}_m}{g^{(\alpha)}_m}\biggr)^2\biggr]\biggr\},
\end{split}
\end{equation}
where
\begin{equation}
  g^{(\alpha)}_m(z) = 
  \begin{cases}
    L^{(\alpha-1)}_m(-z), \quad m=1, 2, 3, \ldots, & \text{for type I}, \\
    L^{(-\alpha-1)}_m(z), \quad m=1, 2, 3, \ldots, \quad \alpha > m-1, & \text{for type II}, \\
    L^{(-\alpha-1)}_m(-z), \quad m=2, 4, 6, \ldots, \quad \alpha > m-1, & \text{for type III}, 
  \end{cases}
\end{equation}
and
\begin{equation}
  z = \frac{1}{2} \omega r^2, \qquad \alpha = l + \frac{1}{2}.
\end{equation}
Type I and type II extended potentials are related to EOP families \cite{cq08a, cq09, odake09, odake10a, odake10b, sasaki, grandati11a, ho11b}, while type III ones are not \cite{cq09, grandati11a}. For the former, the spectrum remains the same as for the conventional potential, in contrast with what happens for the latter. So Eq.\ (\ref{eq:SE-RO}) is changed into
\begin{equation}
  \left(- \frac{d}{dr^2} + V_{l,{\rm ext}}(r)\right) \psi_{l,\nu}(r) = E_{l,\nu} \psi_{l,\nu}(r), \qquad 0 < r < \infty,
  \label{eq:SE-RO-ext}
\end{equation}
where
\begin{equation}
  E_{l,\nu} = 
  \begin{cases}
    \omega \left(2\nu + l + \frac{3}{2}\right), \quad \nu = 0, 1, 2, \ldots, & \text{for type I or II}, \\
    \omega \left(2\nu + l + \frac{7}{2}\right), \quad \nu = -m-1, 0, 1, 2, \ldots, & \text{for type III},
  \end{cases}  \label{eq:RO-ext-spectrum}
\end{equation}
and
\begin{equation}
  \psi_{l,\nu}(r) \propto \frac{\eta_l(z)}{g^{(\alpha)}_m(z)} y^{(\alpha)}_n(z).  \label{eq:RO-ext-wf}
\end{equation}
In (\ref{eq:RO-ext-wf}), $y^{(\alpha)}_n(z)$ denotes a $n$th-degree polynomial in $z$, where $n=m+\nu$, $\nu=0$, 1, 2,~\dots, for type I or II and $n=m+\nu+1$, $\nu=-m-1$, 0, 1, 2,~\dots, for type III.\par
%
%
On performing the PCT
\begin{equation}
  r = e^{- \frac{1}{2}x}, \qquad \psi_{l,\nu}(r) = e^{- \frac{1}{4}x} \varphi_{\nu, A_0}(x)  
\end{equation}
on Eq.\ (\ref{eq:SE-RO-ext}), we get the Schr\"odinger equation  for some extended Morse potential,
\begin{equation}
  \left(- \frac{d^2}{dx^2} + V_{A_{\nu},B,{\rm ext}}(x)\right) \varphi_{\nu,A_0}(x) = \epsilon 
  \varphi_{\nu,A_0}(x),  \label{eq:SE-Morse-ext-map}
\end{equation}
where
\begin{equation}
\begin{split}
  V_{A_{\nu},B,{\rm ext}}(x) & = V_{A_{\nu},B}(x) + V_{A_0,B,{\rm rat}}(x), \\
  V_{A_0,B,{\rm rat}}(x) & = - z \biggl\{\frac{\dot{g}^{(\alpha)}_m}{g^{(\alpha)}_m} + 2z 
       \biggl[\frac{\ddot{g}^{(\alpha)}_m}{g^{(\alpha)}_m}
       - \biggl(\frac{\dot{g}^{(\alpha)}_m}{g^{(\alpha)}_m}\biggr)^2\biggr]\biggr\},
\end{split}  \label{eq:V-Morse-ext-map}
\end{equation}
and
\begin{equation}
\begin{split}
  & A_{\nu} = A_0 + \nu, \qquad A_0 = \frac{\alpha}{2} \text{\ (for type I or II)}, \qquad A_0 = \frac{\alpha}{2}
       + 1 \text{\ (for type III)}, \\
  & B = \frac{1}{4} \omega, \qquad z = 2Be^{-x}, \qquad \epsilon = - \frac{\alpha^2}{4}.
\end{split}
\end{equation}
The eigenfunction $\varphi_{\nu,A_0}(x)$ in (\ref{eq:SE-Morse-ext-map}) can be written as
\begin{equation}
  \varphi_{\nu,A_0}(x) \propto
  \begin{cases}
    \frac{\exp\left(- A_0 x - Be^{-x}\right)}{g^{(2A_0)}_m(z)} y^{(2A_0)}_{m+\nu}(z) & \text{for type I or II}, 
       \\[0.2cm]
    \frac{\exp\left[- (A_0-1) x - Be^{-x}\right]}{g^{(2A_0-2)}_m(z)} y^{(2A_0-2)}_{m+\nu+1}(z) & 
       \text{for type III},
  \end{cases}  \label{eq:Morse-ext-wf-map}
\end{equation}
and corresponds to $\epsilon = - A_0^2$ or $\epsilon = - (A_0-1)^2$, respectively.\par
%
%
As for the conventional potentials, a single rationally-extended radial oscillator potential $V_{l,{\rm ext}}(r)$ is mapped onto a hierarchy of rationally-extended Morse potentials $V_{A_{\nu},B,{\rm ext}}(x)$ with $A_{\nu} = A_0 + \nu$ and $\nu$ running over the range given in (\ref{eq:RO-ext-spectrum}), while $A_0$ and $B$ remain fixed. However, since when $\nu$ is varied, only the core part $V_{A_{\nu},B}(x)$ of the potential is changed and not the whole potential, we cannot deduce other eigenvalues from $\epsilon$ by a reasoning similar to that carried out in Sec.\ 3.1. As a consequence, the potentials (\ref{eq:V-Morse-ext-map}) are QES ones with a single known eigenvalue $\epsilon$ and corresponding eigenfunction $\varphi_{\nu,A_0}(x)$ \cite{turbiner, shifman, ushveridze}. From the known zeros of the polynomials appearing in (\ref{eq:Morse-ext-wf-map}), it is clear that $\varphi_{\nu,A_0}(x)$ is a ground-state wavefunction if $\nu=0$ for type I or II and if $\nu=-m-1$ for type III. Furthermore, if $\nu > 0$ or $\nu \ge 0$, it is a $\nu$th- or $(\nu+1)$th-excited state, respectively.\par
%
%
The results obtained here can be illustrated by considering some special cases. On writing $V_{A_0,B,{\rm rat}}(x) - m$ in a form similar to (\ref{eq:V-rat}), we get
\begin{equation}
\begin{split}
  N_1(x) & = - 3A_0, \\
  N_2(x) & = 2A_0^2, \\
  D(x) & = Be^{-x} + A_0, 
\end{split}
\end{equation}
with $A_0 > 0$ for $m=1$ and type I or II,
\begin{equation}
\begin{split}
  N_1(x) & = -2 (2A_0+1) \bigl(3Be^{-x} + A_0 - 2\bigr), \\
  N_2(x) & = -4 (2A_0+1)^2 \bigl(2Be^{-x} + A_0\bigr), \\
  D(x) & = 2B^2 e^{-2x} + 2 (2A_0+1) Be^{-x} + A_0 (2A_0+1), 
\end{split}
\end{equation}
with $A_0 > 0$ for $m=2$ and type I,
\begin{equation}
\begin{split}
  N_1(x) & = -2 (2A_0-1) \bigl(3Be^{-x} + A_0 + 2\bigr), \\
  N_2(x) & = 4 (2A_0-1)^2 \bigl(2Be^{-x} + A_0\bigr), \\
  D(x) & = 2B^2 e^{-2x} + 2 (2A_0-1) Be^{-x} + A_0 (2A_0-1), 
\end{split}
\end{equation}
with $A_0 > \frac{1}{2}$ for $m=2$ and type II, and
\begin{equation}
\begin{split}
  N_1(x) & = 2 (2A_0-3) \bigl(3Be^{-x} - A_0 - 1\bigr), \\
  N_2(x) & = -4 (2A_0-3)^2 \bigl(2Be^{-x} - A_0 + 1\bigr), \\
  D(x) & = 2B^2 e^{-2x} - 2 (2A_0-3) Be^{-x} + (A_0-1)(2A_0-3), 
\end{split}
\end{equation}
with $A_0 > \frac{3}{2}$ for $m=2$ and type III.\par
%
%
{}For completeness' sake, in Appendix A we review the inverse transformation from the Morse potential to the radial oscillator and apply it to the extended Morse potentials, defined in (\ref{eq:partners}), to generate some QES extended radial oscillators.\par
%
%
\section{Conclusion}

In the present work, we have reconsidered the construction of ES rationally-extended Morse potentials, previously carried out in the framework of the Darboux-B\"acklund transformation \cite{grandati11c} or the prepotential method \cite{ho11c}. On using a first-order SUSYQM approach \cite{cooper} and building on the concept of algebraic deformations of SI potentials \cite{gomez04}, we have obtained the already known family of extended Morse potentials, corresponding to case {\it iii} of SUSYQM. We have called it type III family because it is similar to the L3 series of rationally-extended radial oscillators, for which the polynomial arising in the potential denominator is an $m$th-degree Laguerre polynomial with $m=2$, 4, 6,~\ldots\ and with both negative argument and negative variable.\par
%
%
More importantly, we have pointed out the existence of another family of extensions, corresponding to case {\it ii} of SUSYQM. The members $V^{(m)}_{A,B,{\rm ext}}(x)$ of this family, isospectral to the conventional Morse potential $V_{A+1,B}(x)$, are constructed in terms of $m$th-degree Laguerre polynomials with $m=1$, 2, 3,~\ldots\ and with negative argument, but positive variable, as the L2 series of rationally-extended radial oscillators. For this reason, we have called it type II family. As in the case of extended radial oscillators, the range of parameter values has to be restricted to get singularity-free potentials.\par
%
%
In contrast with what happens for the corresponding extended radial oscillators, however, type II extended Morse potentials are not SI. Nevertheless, they exhibit a kind of `enlarged' SI property in the sense that their partner $V^{(m-1)}_{A-1,B,{\rm ext}}(x)$, in case {\it i} of SUSYQM, is obtained by translating both the parameter $A$ and the polynomial degree $m$, and therefore belongs to the same family of extended potentials.\par
%
%
{}Finally, we have applied the PCT relating the radial oscillator and the Morse potential \cite{cooper, haymaker, de, cq08b} to the rationally-extended radial oscillators belonging to the three known families \cite{cq08a, cq09, odake09, odake10a, odake10b, sasaki, grandati11a, ho11b}. We have shown that to each of the latter potentials, we can associate an infinite hierarchy of extended Morse potentials $V_{A_{\nu},B,{\rm ext}}(x)$, $A_{\nu} = A_0 + \nu$, $\nu=0$, 1, 2,~\ldots, for each one of which one bound state is determined (as in the conventional case). The potentials of this hierarchy contain the same rational extension $V_{A_0,B,{\rm rat}}(x)$, but a different core potential $V_{A_{\nu},B}(x)$, $\nu=0$, 1, 2,~\ldots. As a consequence, the whole spectrum of $V_{A_{\nu},B,{\rm ext}}(x)$ cannot be determined in contrast with what happens in the conventional case. The constructed potentials are therefore QES with a single known bound state \cite{turbiner, shifman, ushveridze}.\par
%
%
Apart from the Morse potential, considered in this paper, and the three potentials mentioned in Sec.\ 1, the Coulomb potential has been rationally extended in two different ways  by using the Darboux-B\"acklund transformation \cite{grandati11c}. Our discovery of type II Morse extensions draws a parallel between the Morse and Coulomb potentials, which previously seemed compromised.\par
%
%
In a future work, we hope to be able to carry out a similar study for those SI potentials associated with Jacobi polynomials and whose solvable rational extensions have not been constructed so far.\par
%
%
\section*{Appendix A. Going from Extended Morse Potentials to Extended Radial Oscillators}

\renewcommand{\theequation}{A.\arabic{equation}}

Let us start from the Schr\"odinger equation for the conventional Morse potential (\ref{eq:Morse-pot}) with eigenvalues $\epsilon^{(A)}_{\nu}$ and wavefunctions $\varphi^{(A)}_{\nu}(x)$, given in (\ref{eq:Morse-spectrum}) and (\ref{eq:Morse-wf}), respectively. Then the map
\begin{equation}
  x = - 2 \log r, \qquad \varphi^{(A)}_{\nu}(x) = r^{-1/2} \psi_{\nu,l_0}(r)  \label{eq:PCT-bis}
\end{equation}
gives rise to the Schr\"odinger equation for a conventional radial oscillator potential
\begin{equation}
  \left(- \frac{d^2}{dr^2} + V_{l_{\nu}}(r)\right) \psi_{\nu,l_0}(r) = E \psi_{\nu,l_0}(r),  \label{eq:SE-RO-map}
\end{equation}
for some fixed energy $E$ defined by
\begin{equation}
  E = \omega \left(l_0 + \frac{3}{2}\right) = 4B (2A+1).
\end{equation}
The potential $V_{l_{\nu}}(r)$ is the same as in (\ref{eq:RO-pot}), but with $l$ replaced by the $\nu$-dependent parameter
\begin{equation}
  l_{\nu} = l_0 - 2\nu, \qquad l_0 = 2A - \frac{1}{2} = \frac{E}{\omega} - \frac{3}{2},
\end{equation}
while the frequency
\begin{equation}
  \omega = 4B
\end{equation}
remains fixed. The wavefunction $\psi_{\nu,l_0}(r)$, associated with the energy $E$, is obtained from (\ref{eq:RO-pot}) and (\ref{eq:PCT-bis}) in the form
\begin{equation}
  \psi_{\nu,l_0}(r) \propto r^{l_{\nu}+1} e^{- \frac{1}{4} \omega r^2} 
  L^{(l_{\nu} + \frac{1}{2})}_{\nu}(z), \qquad z = \frac{1}{2} \omega r^2.
\end{equation}
\par
%
%
The Hamiltonian for a single Morse potential $V_{A,B}(x)$, with given parameters $A$, $B$, is mapped onto a hierarchy of radial oscillator Hamiltonians, corresponding to $V_{l_{\nu}}(r)$ with $l_{\nu} = l_0 - 2\nu$, $\nu=0$, 1, \ldots, $\nu_{\rm max}$, $\nu_{\rm max} = \bigl[\frac{l_0}{2}\bigr]$, and constant $\omega$, $l_0$.\footnote{Note that only some discrete $A$ values lead to $l_0 \in \mathbb{N}$.} Although, for a given $V_{l_{\nu}}(r)$, we get a single eigenvalue $E$ and eigenfunction $\psi_{\nu,l_0}(r)$, the whole spectrum of the radial oscillator potential can be obtained by observing that $V_{l_{\nu}}(r)$ with a given $l_{\nu} = \bar{l}$ may occur in an infinite number of equations of type (\ref{eq:SE-RO-map}), corresponding to $l_0 = \bar{l} + 2\bar{\nu}$, $\bar{\nu}=0$, 1, 2, \ldots, thus leading for $E$ to the well-known expression $\omega \bigl(2\bar{\nu} + \bar{l} + \frac{3}{2}\bigr)$, $\bar{\nu}=0$, 1, 2, \ldots.\par
%
%
Let us now replace the conventional Morse potential by some rationally-extended one, as defined in Eqs.\ (\ref{eq:partners}), (\ref{eq:type-II}), and (\ref{eq:type-III}). We therefore start from the Schr\"odinger equation
\begin{equation}
  \left(- \frac{d^2}{dx^2} + V_{A,B,{\rm ext}}(x)\right) \varphi_{A,\nu}(x) = \epsilon_{A,\nu} \varphi_{A,\nu}(x),
  \qquad - \infty < x < \infty, \label{eq:SE-Morse}
\end{equation}
where
\begin{equation}
  \epsilon_{A,\nu} = 
  \begin{cases}
    - (A+1-\nu)^2, \quad \nu=0, 1, \ldots, \nu_{\rm max}, & \\
        \quad A \le \nu_{\rm max} < A+1, & \text{for type II}, \\
    - (A-1-\nu)^2, \quad \nu=-m-1, 0, 1, \ldots, \nu_{\rm max}, & \\
       \quad A-2 \le \nu_{\rm max} < A-1, & \text{for type III}, 
  \end{cases} \label{eq:Morse-ext-spectrum}
\end{equation}
and
\begin{equation}
  \varphi_{A,\nu}(x) \propto
  \begin{cases}
    \frac{\xi_{A+1}(z)}{g^{(A)}_m(z)} y^{(A)}_n(z), \quad n = m+\nu-1, & \\
    \quad \nu=0, 1, \ldots, \nu_{\rm max}, & \text{for type II}, \\
    \frac{\xi_{A-1}(z)}{g^{(A)}_m(z)} y^{(A)}_n(z), \quad n = m+\nu+1, & \\
    \quad \nu=-m-1, 0, 1, \ldots, \nu_{\rm max}, & \text{for type III},
  \end{cases}
\end{equation}
with $z$ and $\xi_A(z)$ as given in (\ref{eq:z}), while $y^{(A)}_n(z)$ is defined in (\ref{eq:def-y}) or (\ref{eq:def-y-bis}), respectively.\par
%
%
On applying to (\ref{eq:SE-Morse}) a transformation similar to (\ref{eq:PCT-bis}), we get the Schr\"odinger equation for some extended radial oscillator potential,
\begin{equation}
  \left(- \frac{d^2}{dr^2} + V_{l_{\nu},{\rm ext}}(r)\right) \psi_{\nu,l_0}(r) = E \psi_{\nu,l_0}(r),
  \label{eq:SE-RO-ext-map}
\end{equation}
where
\begin{equation}
\begin{split}
  V_{l_{\nu},{\rm ext}}(r) & = V_{l_{\nu}}(r) + V_{l_0,{\rm rat}}(r), \\
  V_{l_0,{\rm rat}}(r) & = - 4\omega\biggl\{\frac{\dot{g}^{(A)}_m}{g^{(A)}_m} + z \biggl[
     \frac{\ddot{g}^{(A)}_m}{g^{(A)}_m} - \biggl(\frac{\dot{g}^{(A)}_m}{g^{(A)}_m}\biggr)^2\biggr]\biggr\},
\end{split}
\end{equation}
and
\begin{equation}
\begin{split}
  l_{\nu} & = l_0 - 2\nu, \qquad l_0 = 2A + \frac{3}{2} \text{\ (for type II}), \qquad l_0 = 2A - \frac{5}{2} 
      \text{\ (for type III}), \\
  \omega & = 4B, \qquad z = \frac{1}{2} \omega r^2, \qquad E = 2\omega \left(A + \frac{1}{2}\right). 
\end{split}
\end{equation}
The eigenfunction $\psi_{\nu,l_0}(r)$ in (\ref{eq:SE-RO-ext-map}) can be written as
\begin{equation}
\begin{split}
  \psi_{\nu,l_0}(r) \propto
  \begin{cases}
    \frac{r^{l_{\nu}+1} \exp(- \frac{1}{4} \omega r^2)}{g^{(\frac{1}{2}(l_0 - \frac{3}{2}))}_m(z)}
        y^{(\frac{1}{2}(l_0 - \frac{3}{2}))}_{m+\nu-1}(z) & \text{for type II}, \\[0.4cm]
    \frac{r^{l_{\nu}+1} \exp(- \frac{1}{4} \omega r^2)}{g^{(\frac{1}{2}(l_0 + \frac{5}{2}))}_m(z)}
        y^{(\frac{1}{2}(l_0 + \frac{5}{2}))}_{m+\nu+1}(z) & \text{for type III},
  \end{cases}
\end{split}
\end{equation}
and corresponds to $E = \omega \bigl(l_0 - \frac{1}{2}\bigr)$ or $E = \omega \bigl(l_0 + \frac{7}{2}\bigr)$, respectively.\par
%
%
Hence, a single rationally-extended Morse potential $V_{A,B,{\rm ext}}(x)$ is mapped onto a hierarchy of rationally-extended radial oscillators $V_{l_{\nu},{\rm ext}}(r)$ with $l_{\nu} = l_0 - 2\nu$ and $\nu$ running over the range given in (\ref{eq:Morse-ext-spectrum}), while $l_0$ and $\omega$ remain fixed. When $\nu$ is varied, only the core part $V_{l_{\nu}}(r)$ of $V_{l_{\nu},{\rm ext}}(r)$ is changed, so that we get a QES potential with a single known eigenvalue $E$ again.\par
%
%
\section*{Acknowledgments}

The author would like to thank Y.\ Grandati for several useful discussions.\par
%
%

\end{document}